\documentclass[ALICE,manyauthors]{cernphprep}
\usepackage{url}
\usepackage{hyperref}

\newcommand {\pt} {\ensuremath p_{\rm T}}

\begin{document}%

\begin{titlepage}

\PHnumber{2013-109}
\PHdate{02 July 2013}

\title{Energy Dependence of the Transverse Momentum Distributions of Charged Particles in pp Collisions Measured by ALICE}

\ShortTitle{Energy Dependence of Charged-Particle $\pt$ Distributions}

\Collaboration{ALICE Collaboration
         \thanks{See Appendix~\ref{app:collab} for the list of collaboration
                      members}}
\ShortAuthor{ALICE Collaboration}

\begin{abstract}
Differential cross sections of charged particles in inelastic pp collisions as a function of $\pt$ have been measured at $\sqrt{s} =$ 0.9, 2.76 and 7 TeV 
at the LHC. The $\pt$ spectra are compared to NLO-pQCD calculations. Though the differential cross section for an individual 
$\sqrt{s}$ cannot be described by NLO-pQCD, the relative increase of cross section with $\sqrt{s}$ is in agreement with NLO-pQCD. Based on these measurements 
and observations, procedures are discussed to construct pp reference spectra at $\sqrt{s} =$ 2.76 and 5.02 TeV up to $\pt$ = 50 GeV/$c$ as required for the 
calculation of the nuclear modification factor in nucleus-nucleus and proton-nucleus collisions.
\end{abstract}
\end{titlepage}
\setcounter{page}{2}

\section{Introduction}

The measurement of charged particle production in proton-proton collisions at high energy gives insight 
into the dynamics of soft and hard interactions. Hard parton-parton scattering processes with large momentum 
transfer are quantitatively described by perturbative Quantum Chromodynamics (pQCD). Measurements at high 
transverse momenta ($p_{\rm T}$) at LHC-energies can help to constrain parton distribution and fragmentation functions in 
current next-to-Leading-Order (NLO) pQCD calculations~\cite{NLO} of charged particle production. As data at various 
$\sqrt{s}$ become available at the LHC, a systematic comparison with current NLO-pQCD calculations over 
a large span of $\sqrt{s}$ is now possible.
However, most particles are produced at low momentum, where particle production is dominated by soft interactions
and only phenomenological approaches can be applied (e.g. PYTHIA~\cite{PYTHIA6}, PHOJET~\cite{PHOJET}) 
to describe the data.
A systematic comparison to data at different values of $\sqrt{s}$ is an essential ingredient to tune these Monte Carlo event 
generators.

Furthermore, the measurement of charged particle transverse momentum spectra in pp collisions serves as a crucial
reference for particle spectra in Pb--Pb collisions. To quantify final state effects due to the creation of a hot and 
dense deconfined matter, commonly referred to as the Quark-Gluon Plasma (QGP), $\pt$ spectra in the two collision systems 
are compared. The observed supression~\cite{RAAfirst} in central Pb--Pb collisions at LHC-energies at high $\pt$ relative to an independent 
superposition of pp collisions is generally attributed to energy loss of the partons as they propagate through the hot and 
dense QCD medium. To enable this comparison a pp reference $\pt$ spectrum at the same $\sqrt{s}$ with the same $\pt$ coverage has
to be provided. Similarly, a pp reference spectrum is also needed for p--Pb collisions to investigate possible 
initial-state effects in the collision.

In this paper we present a measurement of primary charged particle transverse momentum spectra in pp collisions at $\sqrt{s} = 0.9, 2.76$ and $7$ TeV.
Primary charged particles are considered here as all charged particles produced in the collision and their decay products, 
except for particles from weak decays of strange hadrons.
The measurement is performed in the pseudorapidity range $|\eta| <$ 0.8 for particles with $\pt$ $>$ 0.15 GeV/$c$. 
Reference spectra for comparison with Pb--Pb spectra at $\sqrt{s_{\rm NN}}$ = 2.76 TeV and p--Pb spectra at $\sqrt{s_{\rm NN}}$ = 5.02 TeV 
in the corresponding $\pt$ range up to $\pt$ = 50 GeV/$c$ are constructed. 

\section{Experiment and data analysis}

The data were collected by the ALICE apparatus~\cite{ALICE_DET} at the CERN-LHC in 2009--2011. The analysis is based on tracking information from the
Inner Tracking System (ITS) and the Time Projection Chamber (TPC), both located in the central barrel of the experiment.
The minimum-bias interaction trigger was derived using signals from the forward scintillators (VZERO), and the two innermost layers of the ITS, the Silicon Pixel Detector (SPD).   
Details of  the experimental setup used in this analysis are discussed in~\cite{ALICE_PP900}.

The events are selected based on the minimum-bias trigger $\rm{MB_{{OR}}}$ requiring at least one hit in the SPD or VZERO detectors, which are required to be in coincidence with two beam bunches crossing in the ALICE interaction region.
In addition, an offline event selection is applied to reject beam induced (beam-gas, beam-halo) background. The VZERO counters are used to remove these beam-gas 
or beam-halo events by requiring their timing signals to be in coincidence with particles produced in the collision. The background events are also removed by exploiting the correlation between the number of the SPD hits and the number of the SPD tracklets (short track segments reconstructed in the SPD and pointing to the interaction vertex). The beam-gas or beam-halo events typically have a large number of hits in the SPD compared to the number of reconstructed tracklets; this is used to reject background events. In total 6.8~M, 65~M and 150~M pp events at $\sqrt{s}=0.9$, 2.76 and 7 TeV fulfill the $\rm{MB_{{OR}}}$ trigger and offline selection criteria. The typical luminosity for these data taking was about $10^{29}$~${\rm s}^{-1}{\rm cm}^{-2}$. The average number of interactions per bunch crossing varied from 0.05 to 0.1. 

In this analysis the focus is on inelastic (INEL) pp events originating from
single-diffractive, double-diffractive and non-diffractive processes.
The INEL events are selected with an efficiency $\epsilon_{{\rm MB_{OR}}}$ of $91^{+3.2}_{-1.0}\%$, $88.1^{+5.9}_{-3.5}\%$ and $85.2^{+6.2}_{-3.0}\%$ for the three energies.
The trigger efficiencies are determined~\cite{ALICE_CrossSection} based on detector simulations with PYTHIA6~\cite{PYTHIA6} and PHOJET~\cite{PHOJET} event generators.

The primary event vertex is determined based on ITS and TPC information. If no vertex is found using tracks in the ITS and the TPC, it is reconstructed 
from tracklets in the SPD only. Tracks or tracklets are extrapolated to the experimental collision region utilizing the averaged measured beam intersection profile 
in the $x$--$y$ plane perpendicular to the beam axis.

An event is accepted if the $z$-coordinate of the vertex is within \mbox{$\pm 10$~cm} of the center of the interaction region along the beam direction. 
This corresponds to about 1.6 standard deviations from the mean of the reconstructed event vertex distribution for all three energies. 
In this range, the vertex reconstruction efficiency is independent of $z$.
The event vertex reconstruction is fully efficient for events with at least one track in the pseudorapidity range $\left| \eta \right|<1.4$ for all three energies.

Only tracks within a pseudorapidity range of $\left| \eta \right| <$ 0.8 and transverse momenta $p_{\rm T}>0.15$~GeV/$c$ are selected. 
A set of standard cuts based on the number of space points and the quality of the track fit in ITS and TPC is applied to the reconstructed tracks~\cite{ALICE_Second_RAA}.

Efficiency and purity of the primary charged particle selection are estimated using simulations with PYTHIA6~\cite{PYTHIA6} and GEANT3~\cite{GEANT3} for particle
transport and detector response. The overall $p_{\rm T}$-dependent efficiency (tracking efficiency $\times$ acceptance) is 40--73\%,  36--68\% and 
40--73\% at $\sqrt{s}=0.9$,~2.76 and 7~TeV. 
At $\sqrt{s}=2.76$~TeV the overall efficiency is lower than at $\sqrt{s}=0.9$ and 7~TeV due to the smaller number of operational channels in the SPD. 
Contamination of secondary tracks which passed all selection criteria amounts to 7\% at $\pt$$=0.15$~GeV/c and decreases to $\sim$ 0.6\% for $\pt$$>4$~GeV/c. 
In addition, the contribution from secondary tracks originating from weak decays of strange hadrons was scaled up by a factor of 1--1.5 ($p_{\rm{T}}$-dependent) 
to match the contribution in data. The secondary tracks were subtracted bin-by-bin from the $\pt$ spectra.

The $p_{\rm T}$ resolution is estimated from the space point residuals of the track fit. It is verified by the width of the invariant
mass peaks of $\Lambda$, $\overline{\Lambda}$ and K$^0_{\rm{s}}$, reconstructed from their decays into two charged particles.
The relative $p_{\rm T}$ resolution is 3.5\%,~5.5\%~and~9\% at the highest $p_{\rm T}$ of 20,~32~and~50~GeV/$c$
at $\sqrt{s}$~=~0.9,~2.76~and~7~TeV, respectively. From invariant mass distributions $M_{\rm{inv}}(p_{\rm T})$ of $\Lambda$ and K$^0_{\rm{s}}$, the relative uncertainty
on the $p_{\rm T}$ resolution is estimated to be $\approx$20\% for all three energies. To account for the finite $p_{\rm T}$ resolution of tracks, correction factors to the 
spectrum for $p_{\rm T} > 10$~GeV/$c$ are derived using an unfolding procedure. The determination of the correction factors is based on measured tracks without involving simulation. The choice of the unfolding procedure is based on the 
observation that $\pt$ smearing has a small influence on the measured spectrum.  As input to the procedure a power-law parametrization of the measured $p_{\rm T}$ 
spectrum for $p_{\rm T}>10$~GeV/$c$ is used. This parametrization is folded with the $p_{\rm T}$ resolution obtained for a given $\pt$ from the measured track covariance matrix. 
The $p_{\rm T}$ dependent correction factors are extracted from the ratio of the input to the folded parametrization and are applied (bin-by-bin) to the measured $\pt$ spectrum. 
It was checked that the derived correction factors are the same when replacing the measured with the corrected $p_{\rm T}$ distribution in the unfolding procedure. 
The correction factors depend on $\sqrt{s}$ due to the change of the spectral shape and
reach 2\%,~4\%~and~6.5\% at $\sqrt{s}=0.9$,~2.76~and~7~TeV for the highest $p_{\rm {T}}$.
The systematic uncertainty of the momentum scale is $|\Delta(p_{\rm T} )/p_{\rm T}|<0.01$ at $p_{\rm T}=50$~GeV/$c$, as determined from the mass difference
between $\Lambda$ and $\overline{\Lambda}$ and the ratio of positively to negatively charged tracks, assuming charge symmetry at high $p_{\rm T}$.

\begin{table}
\centering
\begin{tabular}{lllll}
$\sqrt{s}$                & 0.9~TeV         & 2.76~TeV        & 7~TeV       \\
\hline
Event vertex selection                & 1.2\%        & 2.3\%        & 0.5\%        \\
Track selection              & 2.5--5.5\%        & 2.3--5.1\%        & 1.9--4.3\%        \\
Tracking efficiency             & 5\%             & 5\%            & 5\%            \\
$p_{\rm T}$~resolution correction     & $<$1.7\%        & $<$1.9\%   & $<$2.6\%        \\
Material budget              & 0.2--1.5\%        &0.2--1.5\%        & 0.2--1.5\%        \\
Particle composition             & 1--2\%        & 1--2\%        & 1--2\%        \\
MC event generator              & 2.5\%            & 2--3\%        & 2--3.5\%        \\
Secondary strange particles        & $<$0.3\%        & $<$0.3\%        & $<$0.3\%        \\
\hline
Total $p_{\rm T}$~dependent        & 6.7--8.2\%        & 6.4--8.0\%    & 6.6--7.9\%        \\
Normalization uncertainty         & +5.1/-4.0\%        & $\pm$1.9\%    & $\pm$3.6\%        \\
\hline
\end{tabular}
\caption{\label{SYSTERRORS} Contribution to the systematic uncertainties on the $p_{\rm T}$~spectra. }
\end{table}

\begin{figure}[t!p]
\begin{center}
	\includegraphics[width=12cm]{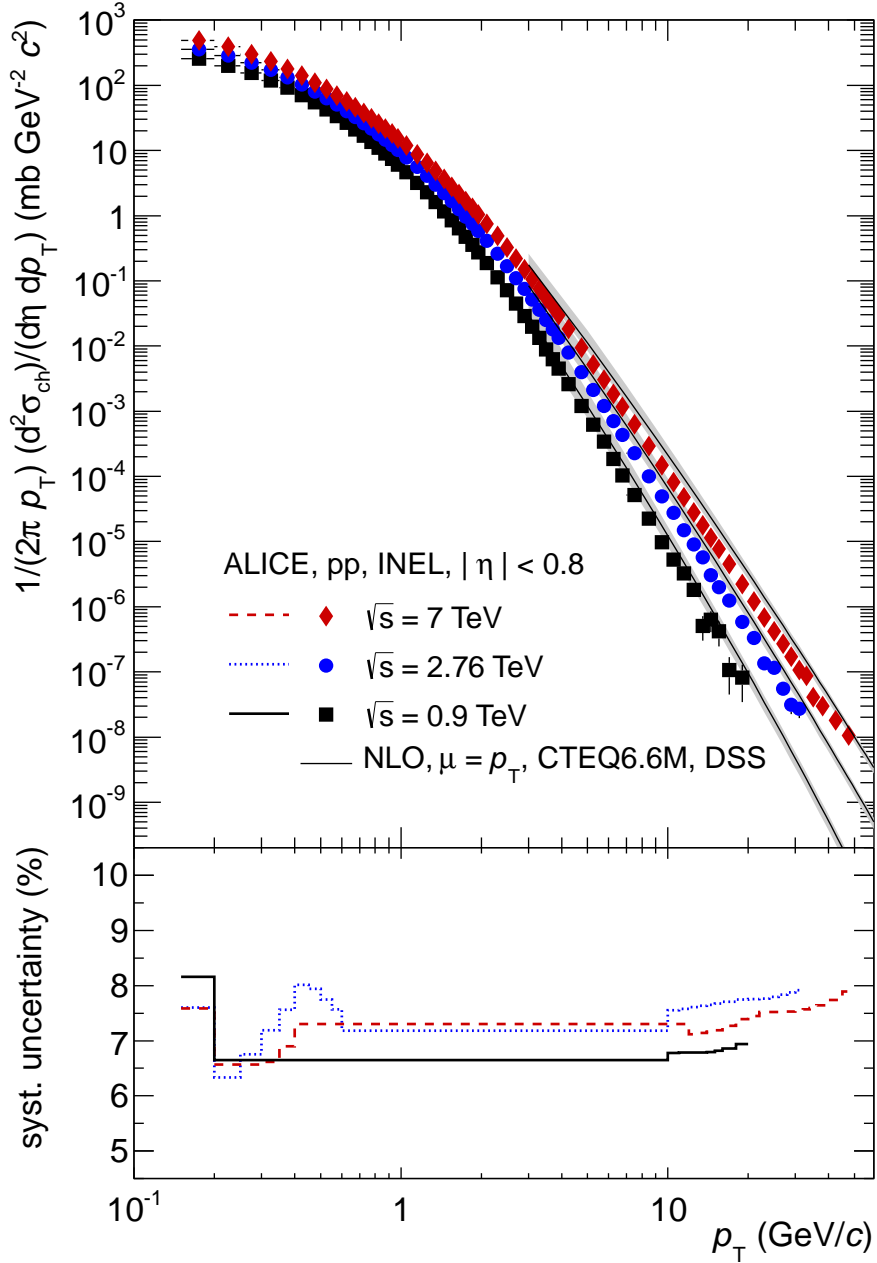}
	\caption{\label{fig:spectraallsyst} (color online) 
         Top: Differential cross section of charged particles in INEL pp collisions at $\sqrt{s} = 0.9$, $2.76$ and $7$ TeV as a function of $\pt$ 
         compared to a NLO-pQCD calculation~\cite{NLO} at the same energy. Only statistical uncertainties are shown.
         Bottom: Systematic uncertainties as a function of $\pt$ for all three energies. 
         The uncertainty on the normalization (compare Table~\ref{SYSTERRORS}) of the spectra is not included.} 
\end{center}
\end{figure}

A summary of the systematic uncertainties is given in Table~\ref{SYSTERRORS}.
The systematic uncertainties on the event selection are determined by changing the lower and upper limits on the $z$-coordinate of the vertex.
Track selection criteria \cite{ALICE_Second_RAA} are varied to determine the corresponding systematic uncertainties resulting in a maximal contribution of 4.3--5.5\% for $p_{\rm T} < 0.6$ GeV/$c$.
The systematic uncertainties on the tracking efficiency are estimated from the difference between data and simulation in the TPC-ITS track matching efficiency.
The systematic uncertainties related to the $p_{\rm T}$ resolution correction are derived from the unfolding procedure including a 
relative uncertainty on the $p_{\rm T}$ resolution, and reach maximum values at the highest $p_{\rm T}$ covered.
The systematic uncertainties on the material budget ($\sim11.5$~\%~$X_0$ \cite{ALICE_Pi0_2011}, where $X_0$ is the radiation length) are estimated by changing the material density (conservatively) by $\pm$10\% 
in the simulation, contributing mostly at $p_{\rm T} < 0.2$ GeV/$c$.
To assess the systematic uncertainties on the tracking efficiency related to the primary particle composition the relative abundance of $\rm{\pi}$, K, p was varied by 30\% in the simulation; 
they contribute mostly at $p_{\rm T} <  0.5$ GeV/$c$.
The Monte Carlo (MC) event generator dependence was studied using PHOJET as a comparison, with the largest contribution at $p_{\rm T} <$ 0.2 GeV/$c$.
The yield of secondary particles from decays of strange hadrons has been varied by 30\% to determine the corresponding uncertainty of maximum 0.3\% at $p_{\rm T} \approx 1$ GeV/$c$.
The total $p_{\rm T}$~dependent systematic uncertainties for the three energies
amount to 6.7--8.2\%, 6.4--8.0\% and 6.6--7.9\% and are shown in the bottom panel of Figure \ref{fig:spectraallsyst}.
They are dominated by the systematic uncertainties on the tracking efficiency. 
There are also comparable contributions related to the track selection ($p_{\rm T} <$ 0.6 GeV/$c$) and $p_{\rm T}$ resolution correction at the highest $p_{\rm T}$ covered.

The systematic uncertainties on the normalization are related to the minimum bias nucleon-nucleon cross section ($\sigma^{{\rm NN}}_{{\rm MB}}$)
determination \cite{ALICE_CrossSection} and amount to +5.1/-4.0\%, $\pm$1.9\% and $\pm$3.6\% for
pp at $\sqrt{s}$~=~0.9~TeV, 2.76~TeV and 7~TeV, respectively.

The differential cross section ${\rm{d}}^2\sigma_{\rm{ch}} / {\rm{d}}\eta {\rm{d}}p_{\rm T}$ is calculated as
${\rm{d}}^2\sigma_{\rm{ch}} / {\rm{d}}\eta {\rm{d}}p_{\rm T} = \sigma_{\rm{MB_{\rm{OR}}}}^{\rm{NN}} \times {\rm{d}}^2 {{N}}^{\rm{MB_{\rm{OR}}}}_{\rm{ch}} / {\rm{d}} \eta {\rm{d}}p_{\rm T}$
with  ${\rm{d}}^2 {{N}}^{\rm{MB_{\rm{OR}}}}_{\rm{ch}} / {\rm{d}} \eta {\rm{d}}p_{\rm T}$ being the per event differential yield of charged particles in minimum bias collisions.
$\sigma_{\rm{MB_{\rm{OR}}}}^{\rm{NN}}$ is determined based on van-der-Meer scans~\cite{ALICE_CrossSection}
as $\sigma_{\rm{MB_{\rm{OR}}}}^{\rm{NN}} = 55.4 \pm 1.0$ ($62.2\pm2.2$)~mb at $\sqrt{s}=2.76$~(7)~TeV.
At $\sqrt{s}=0.9$~TeV van-der-Meer scans were not performed and $\sigma_{\rm{MB_{\rm{OR}}}}^{\rm{NN}}=47.8^{+2.5}_{-3.0}$~mb is
obtained based on detector simulations using the INEL cross section $\sigma^{\rm{NN}}_{\rm{INEL}}=52.5^{+2}_{-3.3}$~mb~\cite{ALICE_CrossSection}. 
$\sigma^{\rm{NN}}_{\rm{INEL}}$ includes the UA5 measurement \cite{UA5_CROSS_SECTION} and re-analysis of the extrapolation to low diffractive masses \cite{ALICE_UA5_REM_CROSS}.

\section{Results}

The differential cross section in INEL pp collisions as a function of $\pt$ is shown in Figure\,\ref{fig:spectraallsyst} for all three measured collision energies. 
At high $\pt$ a clear evolution of the slope from $\sqrt{s}= 0.9$ to 7 TeV can be observed. 
A NLO-pQCD calculation~\cite{NLO} for $\pt$ $>$ 3 GeV/$c$ is compared to the spectra. 
The calculation shows a similar evolution of the high-$\pt$ dependence with $\sqrt{s}$
but overpredicts the data by a factor two~\cite{ALICE_Pi0_2011}~\cite{CMS900}. 
The low systematic uncertainties demonstrate the accuracy of the measurements for all
energies over the full $\pt$ range.

\begin{figure}[t!p]
\begin{center}
	\includegraphics[width=12.0cm]{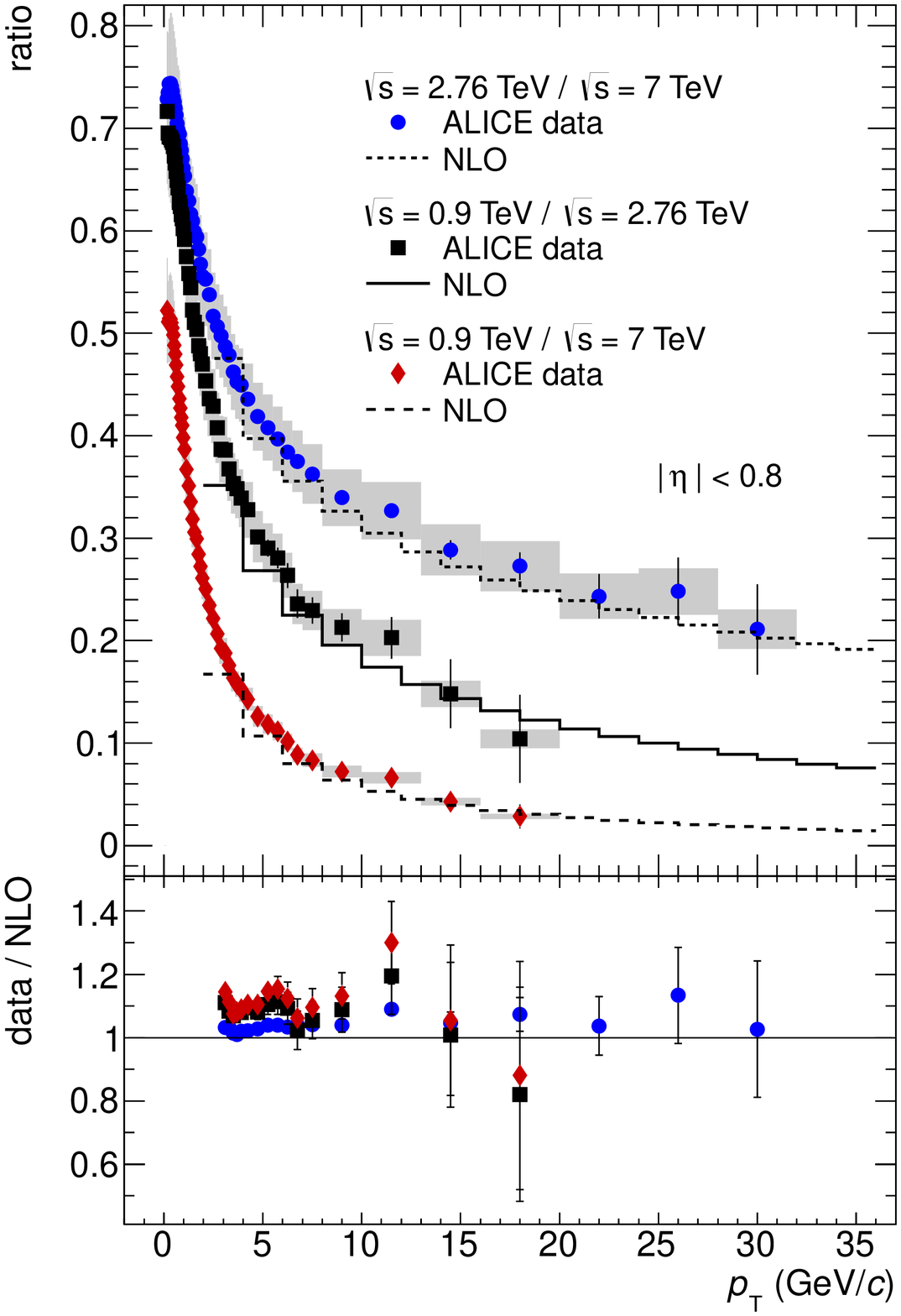}
	\caption{\label{fig:nloratio} (color online) 
         Top: Ratio of differential cross sections of charged particles in INEL pp collisions at different collision energies as a function of $p_{\rm T}$. 
         Grey boxes denote  $p_{\rm T}$ dependent systematic uncertainties. Normalization uncertainties are not shown (see text for details).
         The histograms show the same ratio determined from NLO calculations. Bottom: Ratio of data and NLO calculations  derived from upper panel. 
         A variation of the renormalization and factorization scale of the NLO calculation gives a systematic uncertainty on the double ratio of 0.5--23.6 \% for $0.9$ TeV / $2.76$ TeV, 1.0--37.8\% for 
         $0.9$ TeV / $7$ TeV and 2.4--12.3\% for $2.76$ TeV / $7$ TeV.
}
\end{center}
\end{figure}

Though the $\pt$ dependence of the cross section for a single $\sqrt{s}$ is not well described by NLO-pQCD, 
the relative dependence on $\pt$ of cross sections of two collision energies is described much better.
Figure~\ref{fig:nloratio} shows the ratio between the differential cross section in INEL pp collisions at 
$\sqrt{s} = 2.76$ to $7$ TeV, $0.9$ to $2.76$ TeV and $0.9$ to $7$ TeV as a function of $\pt$ in comparison to 
the same ratio calculated with NLO-pQCD. The total $p_{\rm T}$ dependent systematic uncertainties on the ratios
are evaluated taking into account correlated contributions, and amount to 8.1--9.8\%, 7.8--9.8\% and 
7.9--9.9\% for 0.9~TeV~/~2.76~TeV, 0.9~TeV~/~7~TeV and 2.76~TeV~/~7~TeV. The corresponding normalization 
uncertainties amount to $+5.4\%/-4.4\%$, $+6.2\%/-5.4\%$ and $\pm4.1\%$, and are calculated assuming that 
the normalization uncertainties on the $\pt$ spectra (Table~\ref{SYSTERRORS}) are uncorrelated. 
In all three ratios good agreement between data and NLO-pQCD calculations is found, 
which can be seen in the double ratio of data and NLO-pQCD for the three energy ratios in the lower panel of Figure~\ref{fig:nloratio}.

\section{Construction of a pp reference for $\sqrt{s} = 2.76$ TeV}

For the determination of the nuclear modification factor 
\begin{equation}
R_{\rm AA} (p_{\rm T}) = \frac { {\rm d}^2 {N} _{\rm ch}^{\rm AA}/ {\rm d} \eta {\rm d}p_{\rm T}}
{\langle T_{\rm AA} \rangle \ {\rm d}^2 \sigma_{\rm ch}^{\rm pp} / {\rm d} \eta {\rm d}p_{\rm T}}
\end{equation}
in heavy-ion collisions
a well described pp reference ${\rm d}^2 \sigma_{\rm ch}^{\rm pp} / {\rm d} \eta {\rm d}p_{\rm T}$ at the same center-of-mass energy up to high $\pt$ is essential.
${N}_{\rm ch}^{\rm AA}$ describes the charged particle yield per event in nucleus-nucleus collisions and $\langle T_{\rm AA} \rangle$ is the average nuclear overlap
function~\cite{RAAfirst}~\cite{ALICE_Second_RAA}.
The statistics in the measurement of ${\rm d}^2 \sigma_{\rm ch}^{\rm pp} / {\rm d} \eta {\rm d}p_{\rm T}$ for $\sqrt{s} = 2.76$ TeV reported in this paper allows 
$\pt$ = 32 GeV/$c$ to be reached. 
In order to extrapolate to higher $\pt$, the measured cross section needs to be parametrized.

As can be seen in Figure~\ref{fig:spectraallsyst} for $\pt$ $>$ 10 GeV/$c$ the pp spectrum at $\sqrt{s} = 2.76$ TeV shows a clear power-law 
dependence on $p_{\rm T}$.
To constrain the parametrization better by including data points at lower $p_{\rm T}$,  
${\rm d}^2 \sigma_{\rm ch}^{\rm pp} / {\rm d} \eta {\rm d}p_{\rm T}$ has been parametrized by a so-called modified Hagedorn function~\cite{Hagedorn}
\begin{equation}
	\frac{1}{2\pi p_{\rm T}}\frac{{\rm d}^2 \sigma_{\rm ch}^{\rm pp}}{{\rm d} \eta {\rm d}p_{\rm T}} = A \frac{p_{\rm T}}{m_{\rm T}} \left (1+ \frac{p_{\rm T}}{p_{\rm T,0}}\right)^{-n}
	\label{eq:hagedorn} 
\end{equation}
where $m_{\rm T}$ denotes the transverse mass $m_{\rm T} = \sqrt{m_0^2 + p_{\rm T}^2}$, with $m_0=140$ MeV/$c$ assumed for 
all tracks. For small $p_{\rm T}$, the term $\left (1+ \frac{p_{\rm T}}{p_{\rm T,0}}\right)^{-n}$ behaves like an exponential function with an inverse 
slope parameter of $p_{\rm T,0}/n$ while for large $\pt$ the Hagedorn function behaves like a power-law function. 

To determine the extrapolation to high $p_{\rm T}$, ${\rm d}^2 \sigma_{\rm ch}^{\rm pp} / {\rm d} \eta {\rm d}p_{\rm T}$ is parametrized for $\pt$ $>$ 5 GeV/$c$.
For 5~GeV/$c$~$<$~$\pt$~$<$~10~GeV/$c$ the exponential part of the Hagedorn function acts as a correction term to the power-law part in the function. \\

\begin{figure}[t!p]
\begin{center}
	\includegraphics[width=12.0cm]{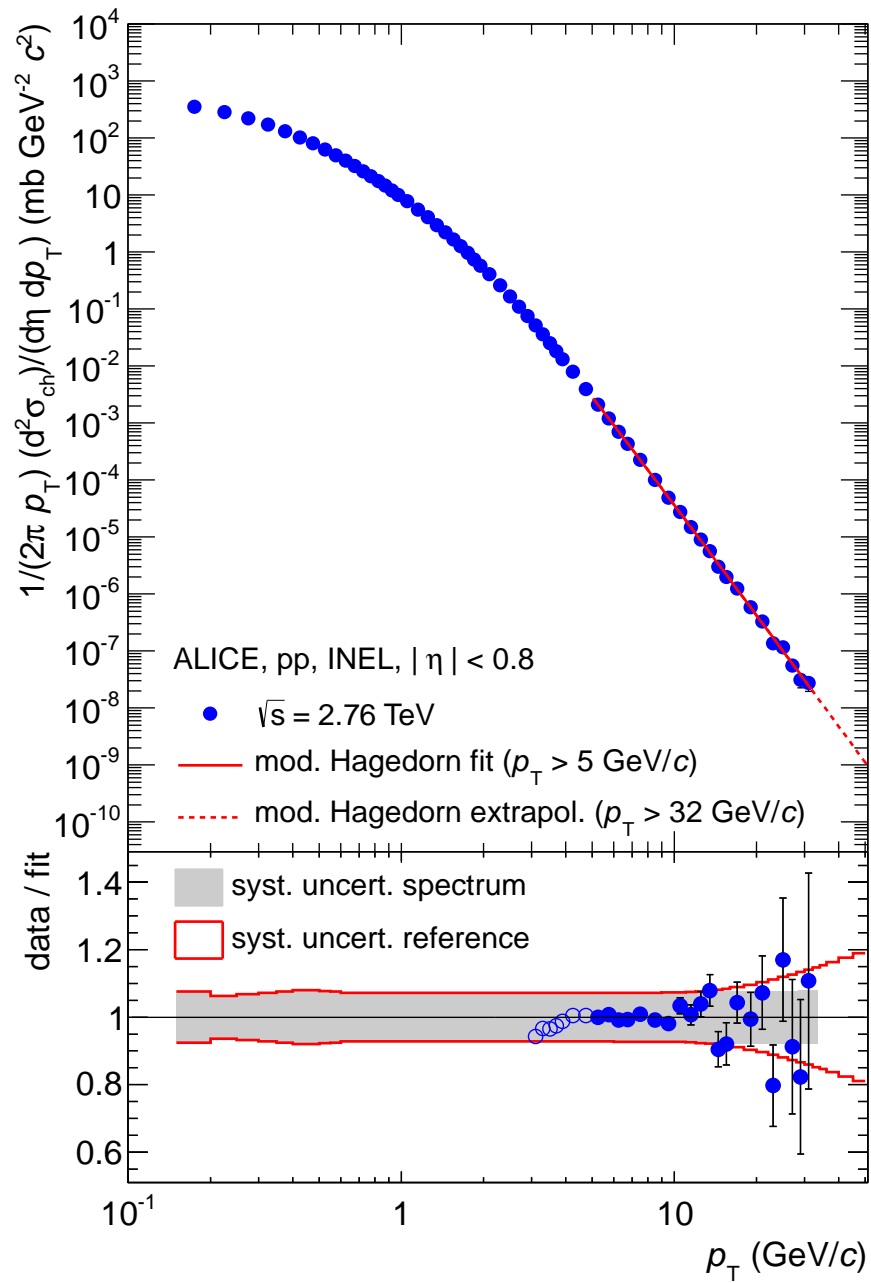}
	\caption{\label{fig:reference}  (color online) 
          Top: Differential cross section of charged particles in INEL pp collisions at $\sqrt{s} = 2.76$ TeV as a function of $\pt$ 
          together with the parametrization ($\pt$ $>$ 5 GeV/$c$) described in the text. 
          Bottom: Ratio of data to parametrization. The grey band indicates the total $p_{\rm T}$ dependent systematic uncertainty of the 
          data, open circles show data points only used for the evaluation of the systematic uncertainty of the parametrization.}
\end{center}
\end{figure}

Figure~\ref{fig:reference} shows the differential cross section in INEL pp collisions as a function of $\pt$ for $\sqrt{s} = 2.76$~TeV
together with the parametrization for $\pt$ $>$ 5 GeV/$c$. 
The ratio between data and parametrization in the lower panel demonstrates the good agreement
of the parametrization with the data.
The grey band indicates the total $p_{\rm T}$ dependent systematic uncertainty of the measured spectrum as presented in Table~\ref{SYSTERRORS}.

To estimate the systematic uncertainty of the parametrization and
extrapolation, the lower boundary of the fit range of the Hagedorn
parametrization is varied between $p_{\rm T} = 3$ GeV/$c$ and 
$p_{\rm T} = 7$ GeV/$c$, while the upper boundary is fixed to the highest data
point measured at $p_{\rm T} = 32$ GeV/$c$.
Together with the systematic uncertainties on the measured differential cross section as shown in 
Table~\ref{SYSTERRORS} this results in a total systematic uncertainty on the reference at 
$\sqrt{s} = 2.76$ TeV of 6.4\% for low $\pt$ up to 19\% at $\pt$~=~50~GeV/$c$.

The final pp reference for the determination of $R_{\rm AA}$ at $\sqrt{s} = 2.76$ TeV is constructed
from the measured data points up to $\pt$ = 5 GeV/$c$ and the parametrization for $\pt$ $>$ 5 GeV/$c$. 
Statistical uncertainties in the extrapolated part of the reference are obtained from the covariance matrix
of the parametrization. The systematic uncertainties on the spectrum are propagated to the reference 
by application of the full extrapolation procedure using the measured data points shifted up and down 
by the total systematic uncertainty.

\begin{figure}[t!p]
\begin{center}
       \includegraphics[width=12.0cm]{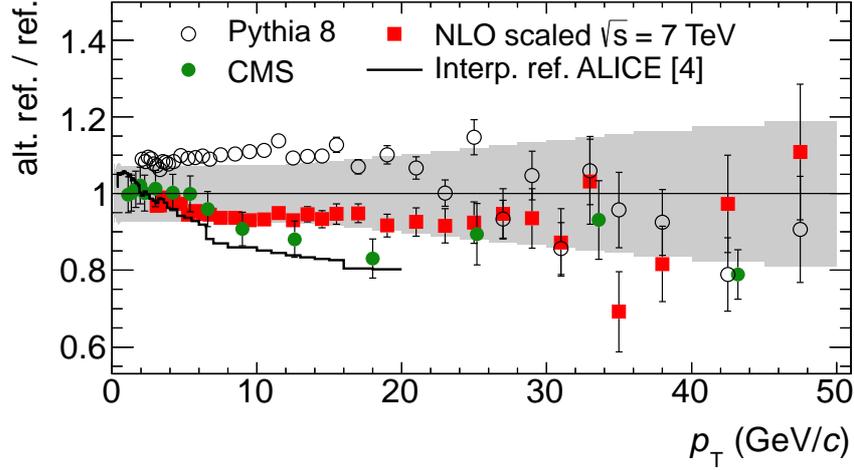}
       \caption{\label{fig:refratio}  (color online)  Ratio of alternative references to the new constructed pp reference at $\sqrt{s} = 2.76$ TeV as discussed in the text.
         The grey band indicates the total $p_{\rm T}$ dependent systematic uncertainty as discussed in the text. 
         The overall normalization systematic uncertainties $\pm1.9\%$ ($\pm6\%$) for ALICE (CMS) are not shown.}
\end{center}
\end{figure}

This reference is compared to alternative measurements and approaches. Figure~\ref{fig:refratio} shows the ratio between
alternative pp references and the reference at $\sqrt{s} = 2.76$ TeV presented in this paper.
Above $\pt$$=20$~GeV/$c$, all references agree within the systematic uncertainties.
Simulations with the PYTHIA8 generator~\cite{PYTHIA8} agree with the new reference for $\pt$ $>$ 15 GeV/$c$.
Below $\pt$$=20$~GeV/$c$, the shape of the PYTHIA8 spectrum is similar to the measured reference.
A pp reference presented by the CMS collaboration~\cite{CMS} agrees best for $\pt$ $<$ 6 GeV/$c$.  
The overall normalization systematic uncertainties $\pm1.9\%$ ($\pm6\%$) for ALICE (CMS) are not included in the comparison.
A reference based on an interpolation between measured yields at $\sqrt{s} = 0.9$ and $7$ TeV
as discussed in~\cite{RAAfirst} does not agree with the new reference for $\pt$ $>$ 6 GeV/$c$.
Finally a scaling of the measured differential cross section in INEL pp collisions at $\sqrt{s} = 7$ TeV
with the ratio of pQCD calculations (as shown in Figure~\ref{fig:nloratio})

\begin{equation}
{\rm d}^2 \sigma_{\rm ch}^{\rm pp} / {\rm d} \eta {\rm d}p_{\rm T} \mid_{\rm 2.76 TeV}
=
\frac{
{\rm d}^2 \sigma_{\rm ch}^{\rm pp} / {\rm d} \eta {\rm d}p_{\rm T} \mid_{\rm NLO, 2.76 TeV}
}
{
{\rm d}^2 \sigma_{\rm ch}^{\rm pp} / {\rm d} \eta {\rm d}p_{\rm T} \mid_{\rm NLO, 7 TeV}
}
\times
{\rm d}^2 \sigma_{\rm ch}^{\rm pp} / {\rm d} \eta {\rm d}p_{\rm T} \mid_{\rm 7 TeV}
\label{eq:NLO} 
\end{equation}

agrees well in shape and normalization with the measured data over a wide range in $p_{\rm T}$. 
The systematic uncertainty of the new reference is indicated in Figure~\ref{fig:refratio} as a grey band for
comparison.

\begin{figure}[t!p]
\begin{center}
	\includegraphics[width=12.0cm]{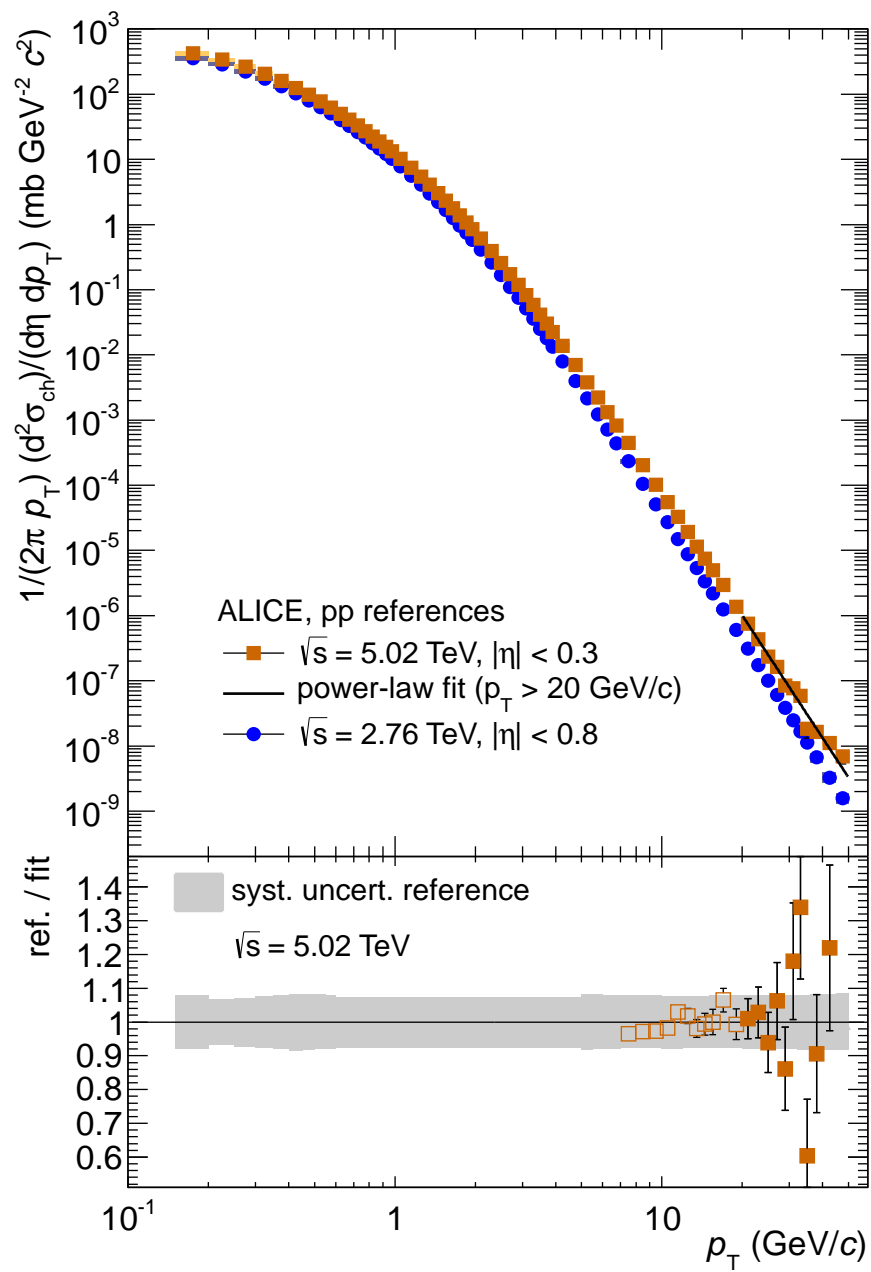}
	\caption{\label{fig:references}  (color online) Top: 
          Constructed pp references for $\sqrt{s}$ = 2.76 and $\sqrt{s}$ = 5.02 TeV. 
          Bottom: Comparison of NLO-scaled reference and parametrization. The parametrization is used for $\pt$ $>$ 20 GeV/$c$.
          The grey band indicates the total $p_{\rm T}$ dependent systematic uncertainty as discussed in the text.}
\end{center}
\end{figure}

\section{Construction of a pp reference for $\sqrt{s} = 5.02$ TeV}

Similar to $R_{\rm AA}$, a nuclear modification factor $R_{\rm pA}$  in proton-lead collisions has been studied~\cite{RpA} at 
$\sqrt{s} = 5.02$ TeV. No measured pp reference is available at this collision energy.
Due to the asymmetric p-Pb collision system, the $\eta$ coverage of the detector is shifted with respect to the symmetric pp or Pb--Pb collisions. 
To obtain a maximum overlap between the pp and p-Pb systems, a pp reference is needed for $|\eta| < 0.3$. 
To construct the pp reference at this energy, different methods for three $\pt$-ranges are combined.\\
0.15 $<$ $\pt$ $<$ 5 GeV/$c$:
As NLO-pQCD becomes unreliable for small $p_{\rm T}$, the measured differential cross sections for pp collisions of $\sqrt{s}$ =  2.76 and 7 TeV 
are interpolated for a given $p_{\rm T}$, assuming a power-law behaviour of the $\sqrt{s}$ dependence of the cross section. 
Here the maximum relative systematic uncertainty of the underlying measurements has been assigned as systematic uncertainty.\\ 
5 $<$ $\pt$ $<$ 20 GeV/$c$: 
The measured differential cross section for pp collisions at $\sqrt{s} = 7$ TeV is scaled to $\sqrt{s} = 5.02$ TeV  
using the NLO-pQCD calculations (Equation~\ref{eq:NLO}). Systematic uncertainties are determined
by taking into account differences to an interpolated reference as well as to a scaled reference using $\mu = p_{\rm T}/2$ and $\mu = 2p_{\rm T}$
as alternative choices for the renormalization and factorization scales.\\
$\pt$ $>$ 20 GeV/$c$: 
The NLO-scaled reference is parametrized in the range 20 $<$ $\pt$ $<$ 50 GeV/$c$ by a power-law function and
the parametrization is used.

The constructed pp reference for $\sqrt{s} = 5.02$ TeV is shown in Figure~\ref{fig:references} together with the reference
for $\sqrt{s}$~=~2.76~TeV discussed above. For $\pt$ $>$ 20 GeV/$c$ the data points show the NLO-scaled reference which is parametrized
by a power-law function (line) to obtain the final reference at $\sqrt{s} = 5.02$ TeV. 
In the bottom part of the figure a comparison of the NLO-scaled reference and the parametrization is shown.

\section{Summary}

Differential cross sections of charged particles in inelastic pp collisions as a function of $\pt$ have been presented for $\sqrt{s} =$ 0.9, 2.76 and 7 TeV.
Comparisons of the $\pt$ spectra with NLO-pQCD calculations show that the cross section for an individual value of $\sqrt{s}$ cannot be described by the calculation.
The relative increase of cross section with $\sqrt{s}$ is well described by NLO-pQCD, however. The systematic comparison of the energy dependence can help
to tune the model dependent ingredients in the calculation.
Utilizing these observations and measurements procedures are discussed to construct pp reference spectra at $\sqrt{s} =$ 2.76 ($|\eta|<0.8$) and 5.02 TeV ($|\eta|<0.3$) 
in the corresponding $\pt$ range of charged particle $\pt$ spectra in Pb--Pb and p--Pb collisions measured by the ALICE experiment. 
The reference spectra are used for the calculation of the nuclear modification factors $R_{\rm AA}$~\cite{ALICE_Second_RAA} and $R_{\rm pA}$~\cite{RpA}. The systematic uncertainties related to the pp reference were significantly reduced with respect to the previous measurement by using the $\pt$ distribution measured in pp collisions at $\sqrt{s}$~=~2.76~TeV. 

\newenvironment{acknowledgement}{\relax}{\relax}
\begin{acknowledgement}
\section*{Acknowledgements}
The ALICE collaboration would like to thank all its engineers and technicians for their invaluable contributions to the construction of the experiment and the CERN accelerator teams for the outstanding performance of the LHC complex.
The ALICE collaboration acknowledges the following funding agencies for their support in building and
running the ALICE detector:
State Committee of Science,  World Federation of Scientists (WFS)
and Swiss Fonds Kidagan, Armenia,
Conselho Nacional de Desenvolvimento Cient\'{\i}fico e Tecnol\'{o}gico (CNPq), Financiadora de Estudos e Projetos (FINEP),
Funda\c{c}\~{a}o de Amparo \`{a} Pesquisa do Estado de S\~{a}o Paulo (FAPESP);
National Natural Science Foundation of China (NSFC), the Chinese Ministry of Education (CMOE)
and the Ministry of Science and Technology of China (MSTC);
Ministry of Education and Youth of the Czech Republic;
Danish Natural Science Research Council, the Carlsberg Foundation and the Danish National Research Foundation;
The European Research Council under the European Community's Seventh Framework Programme;
Helsinki Institute of Physics and the Academy of Finland;
French CNRS-IN2P3, the `Region Pays de Loire', `Region Alsace', `Region Auvergne' and CEA, France;
German BMBF and the Helmholtz Association;
General Secretariat for Research and Technology, Ministry of
Development, Greece;
Hungarian OTKA and National Office for Research and Technology (NKTH);
Department of Atomic Energy and Department of Science and Technology of the Government of India;
Istituto Nazionale di Fisica Nucleare (INFN) and Centro Fermi -
Museo Storico della Fisica e Centro Studi e Ricerche "Enrico
Fermi", Italy;
MEXT Grant-in-Aid for Specially Promoted Research, Ja\-pan;
Joint Institute for Nuclear Research, Dubna;
National Research Foundation of Korea (NRF);
CONACYT, DGAPA, M\'{e}xico, ALFA-EC and the EPLANET Program
(European Particle Physics Latin American Network)
Stichting voor Fundamenteel Onderzoek der Materie (FOM) and the Nederlandse Organisatie voor Wetenschappelijk Onderzoek (NWO), Netherlands;
Research Council of Norway (NFR);
Polish Ministry of Science and Higher Education;
National Authority for Scientific Research - NASR (Autoritatea Na\c{t}ional\u{a} pentru Cercetare \c{S}tiin\c{t}ific\u{a} - ANCS);
Ministry of Education and Science of Russian Federation, Russian
Academy of Sciences, Russian Federal Agency of Atomic Energy,
Russian Federal Agency for Science and Innovations and The Russian
Foundation for Basic Research;
Ministry of Education of Slovakia;
Department of Science and Technology, South Africa;
CIEMAT, EELA, Ministerio de Econom\'{i}a y Competitividad (MINECO) of Spain, Xunta de Galicia (Conseller\'{\i}a de Educaci\'{o}n),
CEA\-DEN, Cubaenerg\'{\i}a, Cuba, and IAEA (International Atomic Energy Agency);
Swedish Research Council (VR) and Knut $\&$ Alice Wallenberg
Foundation (KAW);
Ukraine Ministry of Education and Science;
United Kingdom Science and Technology Facilities Council (STFC);
The United States Department of Energy, the United States National
Science Foundation, the State of Texas, and the State of Ohio.

\end{acknowledgement}

\newpage

\appendix
\section{The ALICE Collaboration}
\label{app:collab}



\begingroup
\small
\begin{flushleft}
B.~Abelev\Irefn{org69}\And
J.~Adam\Irefn{org36}\And
D.~Adamov\'{a}\Irefn{org77}\And
A.M.~Adare\Irefn{org124}\And
M.M.~Aggarwal\Irefn{org81}\And
G.~Aglieri~Rinella\Irefn{org33}\And
M.~Agnello\Irefn{org104}\textsuperscript{,}\Irefn{org87}\And
A.G.~Agocs\Irefn{org123}\And
A.~Agostinelli\Irefn{org25}\And
Z.~Ahammed\Irefn{org119}
% ...
\renewcommand\labelenumi{\textsuperscript{\theenumi}~}

\section*{Affiliation notes}
\renewcommand\theenumi{\roman{enumi}}
\begin{Authlist}
\item \Adef{0}Deceased
\item \Adef{idp3702560}{Also at: M.V.Lomonosov Moscow State University, D.V.Skobeltsyn Institute of Nuclear Physics, Moscow, Russia}
\item \Adef{idp3946960}{Also at: University of Belgrade, Faculty of Physics and "Vin\v{c}a" Institute of Nuclear Sciences, Belgrade, Serbia}
\item \Adef{idp4773664}{Also at: Institute of Theoretical Physics, University of Wroclaw, Wroclaw, Poland}
\end{Authlist}

\section*{Collaboration Institutes}
\renewcommand\theenumi{\arabic{enumi}~}
\begin{Authlist}

\item \Idef{org1}A. I. Alikhanyan National Science Laboratory (Yerevan Physics Institute) Foundation, Yerevan, Armenia
\item \Idef{org2}Benem\'{e}rita Universidad Aut\'{o}noma de Puebla, Puebla, Mexico
\item \Idef{org3}Bogolyubov Institute for Theoretical Physics, Kiev, Ukraine
\item \Idef{org4}Budker Institute for Nuclear Physics, Novosibirsk, Russia
\item \Idef{org5}California Polytechnic State University, San Luis Obispo, California, United States
\item \Idef{org6}Central China Normal University, Wuhan, China
\item \Idef{org7}Centre de Calcul de l'IN2P3, Villeurbanne, France 
\item \Idef{org8}Centro de Aplicaciones Tecnol\'{o}gicas y Desarrollo Nuclear (CEADEN), Havana, Cuba
\item \Idef{org9}Centro de Investigaciones Energ\'{e}ticas Medioambientales y Tecnol\'{o}gicas (CIEMAT), Madrid, Spain
\item \Idef{org10}Centro de Investigaci\'{o}n y de Estudios Avanzados (CINVESTAV), Mexico City and M\'{e}rida, Mexico
\item \Idef{org11}Centro Fermi - Museo Storico della Fisica e Centro Studi e Ricerche ``Enrico Fermi'', Rome, Italy
\item \Idef{org12}Chicago State University, Chicago, United States
\item \Idef{org13}Commissariat \`{a} l'Energie Atomique, IRFU, Saclay, France
\item \Idef{org14}COMSATS Institute of Information Technology (CIIT), Islamabad, Pakistan
\item \Idef{org15}Departamento de F\'{\i}sica de Part\'{\i}culas and IGFAE, Universidad de Santiago de Compostela, Santiago de Compostela, Spain
\item \Idef{org16}Department of Physics Aligarh Muslim University, Aligarh, India
\item \Idef{org17}Department of Physics and Technology, University of Bergen, Bergen, Norway
\item \Idef{org18}Department of Physics, Ohio State University, Columbus, Ohio, United States
\item \Idef{org19}Department of Physics, Sejong University, Seoul, South Korea
\item \Idef{org20}Department of Physics, University of Oslo, Oslo, Norway
\item \Idef{org21}Dipartimento di Fisica dell'Universit\`{a} and Sezione INFN, Cagliari, Italy
\item \Idef{org22}Dipartimento di Fisica dell'Universit\`{a} and Sezione INFN, Trieste, Italy
\item \Idef{org23}Dipartimento di Fisica dell'Universit\`{a} and Sezione INFN, Turin, Italy
\item \Idef{org24}Dipartimento di Fisica dell'Universit\`{a} `La Sapienza` and Sezione INFN, Rome, Italy
\item \Idef{org25}Dipartimento di Fisica e Astronomia dell'Universit\`{a} and Sezione INFN, Bologna, Italy
\item \Idef{org26}Dipartimento di Fisica e Astronomia dell'Universit\`{a} and Sezione INFN, Catania, Italy
\item \Idef{org27}Dipartimento di Fisica e Astronomia dell'Universit\`{a} and Sezione INFN, Padova, Italy
\item \Idef{org28}Dipartimento di Fisica `E.R.~Caianiello' dell'Universit\`{a} and Gruppo Collegato INFN, Salerno, Italy
\item \Idef{org29}Dipartimento di Scienze e Innovazione Tecnologica dell'Universit\`{a} del Piemonte Orientale and Gruppo Collegato INFN, Alessandria, Italy
\item \Idef{org30}Dipartimento Interateneo di Fisica `M.~Merlin' and Sezione INFN, Bari, Italy
\item \Idef{org31}Division of Experimental High Energy Physics, University of Lund, Lund, Sweden
\item \Idef{org32}Eberhard Karls Universit\"{a}t T\"{u}bingen, T\"{u}bingen, Germany
\item \Idef{org33}European Organization for Nuclear Research (CERN), Geneva, Switzerland
\item \Idef{org34}Faculty of Engineering, Bergen University College, Bergen, Norway
\item \Idef{org35}Faculty of Mathematics, Physics and Informatics, Comenius University, Bratislava, Slovakia
\item \Idef{org36}Faculty of Nuclear Sciences and Physical Engineering, Czech Technical University in Prague, Prague, Czech Republic
\item \Idef{org37}Faculty of Science, P.J.~\v{S}af\'{a}rik University, Ko\v{s}ice, Slovakia
\item \Idef{org38}Frankfurt Institute for Advanced Studies, Johann Wolfgang Goethe-Universit\"{a}t Frankfurt, Frankfurt, Germany
\item \Idef{org39}Gangneung-Wonju National University, Gangneung, South Korea
\item \Idef{org40}Helsinki Institute of Physics (HIP), Helsinki, Finland
\item \Idef{org41}Hiroshima University, Hiroshima, Japan
\item \Idef{org42}Indian Institute of Technology Bombay (IIT), Mumbai, India
\item \Idef{org43}Indian Institute of Technology Indore, India (IITI)
\item \Idef{org44}Institut de Physique Nucl\'{e}aire d'Orsay (IPNO), Universit\'{e} Paris-Sud, CNRS-IN2P3, Orsay, France
\item \Idef{org45}Institut f\"{u}r Informatik, Johann Wolfgang Goethe-Universit\"{a}t Frankfurt, Frankfurt, Germany
\item \Idef{org46}Institut f\"{u}r Kernphysik, Johann Wolfgang Goethe-Universit\"{a}t Frankfurt, Frankfurt, Germany
\item \Idef{org47}Institut f\"{u}r Kernphysik, Technische Universit\"{a}t Darmstadt, Darmstadt, Germany
\item \Idef{org48}Institut f\"{u}r Kernphysik, Westf\"{a}lische Wilhelms-Universit\"{a}t M\"{u}nster, M\"{u}nster, Germany
\item \Idef{org49}Institut Pluridisciplinaire Hubert Curien (IPHC), Universit\'{e} de Strasbourg, CNRS-IN2P3, Strasbourg, France
\item \Idef{org50}Institute for High Energy Physics, Protvino, Russia
\item \Idef{org51}Institute for Nuclear Research, Academy of Sciences, Moscow, Russia
\item \Idef{org52}Institute for Subatomic Physics of Utrecht University, Utrecht, Netherlands
\item \Idef{org53}Institute for Theoretical and Experimental Physics, Moscow, Russia
\item \Idef{org54}Institute of Experimental Physics, Slovak Academy of Sciences, Ko\v{s}ice, Slovakia
\item \Idef{org55}Institute of Physics, Academy of Sciences of the Czech Republic, Prague, Czech Republic
\item \Idef{org56}Institute of Physics, Bhubaneswar, India
\item \Idef{org57}Institute of Space Science (ISS), Bucharest, Romania
\item \Idef{org58}Instituto de Ciencias Nucleares, Universidad Nacional Aut\'{o}noma de M\'{e}xico, Mexico City, Mexico
\item \Idef{org59}Instituto de F\'{\i}sica, Universidad Nacional Aut\'{o}noma de M\'{e}xico, Mexico City, Mexico
\item \Idef{org60}iThemba LABS, National Research Foundation, Somerset West, South Africa
\item \Idef{org61}Joint Institute for Nuclear Research (JINR), Dubna, Russia
\item \Idef{org62}Korea Institute of Science and Technology Information, Daejeon, South Korea
\item \Idef{org63}KTO Karatay University, Konya, Turkey
\item \Idef{org64}Laboratoire de Physique Corpusculaire (LPC), Clermont Universit\'{e}, Universit\'{e} Blaise Pascal, CNRS--IN2P3, Clermont-Ferrand, France
\item \Idef{org65}Laboratoire de Physique Subatomique et de Cosmologie (LPSC), Universit\'{e} Joseph Fourier, CNRS-IN2P3, Institut Polytechnique de Grenoble, Grenoble, France
\item \Idef{org66}Laboratori Nazionali di Frascati, INFN, Frascati, Italy
\item \Idef{org67}Laboratori Nazionali di Legnaro, INFN, Legnaro, Italy
\item \Idef{org68}Lawrence Berkeley National Laboratory, Berkeley, California, United States
\item \Idef{org69}Lawrence Livermore National Laboratory, Livermore, California, United States
\item \Idef{org70}Moscow Engineering Physics Institute, Moscow, Russia
\item \Idef{org71}National Centre for Nuclear Studies, Warsaw, Poland
\item \Idef{org72}National Institute for Physics and Nuclear Engineering, Bucharest, Romania
\item \Idef{org73}National Institute of Science Education and Research, Bhubaneswar, India
\item \Idef{org74}Niels Bohr Institute, University of Copenhagen, Copenhagen, Denmark
\item \Idef{org75}Nikhef, National Institute for Subatomic Physics, Amsterdam, Netherlands
\item \Idef{org76}Nuclear Physics Group, STFC Daresbury Laboratory, Daresbury, United Kingdom
\item \Idef{org77}Nuclear Physics Institute, Academy of Sciences of the Czech Republic, \v{R}e\v{z} u Prahy, Czech Republic
\item \Idef{org78}Oak Ridge National Laboratory, Oak Ridge, Tennessee, United States
\item \Idef{org79}Petersburg Nuclear Physics Institute, Gatchina, Russia
\item \Idef{org80}Physics Department, Creighton University, Omaha, Nebraska, United States
\item \Idef{org81}Physics Department, Panjab University, Chandigarh, India
\item \Idef{org82}Physics Department, University of Athens, Athens, Greece
\item \Idef{org83}Physics Department, University of Cape Town, Cape Town, South Africa
\item \Idef{org84}Physics Department, University of Jammu, Jammu, India
\item \Idef{org85}Physics Department, University of Rajasthan, Jaipur, India
\item \Idef{org86}Physikalisches Institut, Ruprecht-Karls-Universit\"{a}t Heidelberg, Heidelberg, Germany
\item \Idef{org87}Politecnico di Torino, Turin, Italy
\item \Idef{org88}Purdue University, West Lafayette, Indiana, United States
\item \Idef{org89}Pusan National University, Pusan, South Korea
\item \Idef{org90}Research Division and ExtreMe Matter Institute EMMI, GSI Helmholtzzentrum f\"ur Schwerionenforschung, Darmstadt, Germany
\item \Idef{org91}Rudjer Bo\v{s}kovi\'{c} Institute, Zagreb, Croatia
\item \Idef{org92}Russian Federal Nuclear Center (VNIIEF), Sarov, Russia
\item \Idef{org93}Russian Research Centre Kurchatov Institute, Moscow, Russia
\item \Idef{org94}Saha Institute of Nuclear Physics, Kolkata, India
\item \Idef{org95}School of Physics and Astronomy, University of Birmingham, Birmingham, United Kingdom
\item \Idef{org96}Secci\'{o}n F\'{\i}sica, Departamento de Ciencias, Pontificia Universidad Cat\'{o}lica del Per\'{u}, Lima, Peru
\item \Idef{org97}Sezione INFN, Bari, Italy
\item \Idef{org98}Sezione INFN, Bologna, Italy
\item \Idef{org99}Sezione INFN, Cagliari, Italy
\item \Idef{org100}Sezione INFN, Catania, Italy
\item \Idef{org101}Sezione INFN, Padova, Italy
\item \Idef{org102}Sezione INFN, Rome, Italy
\item \Idef{org103}Sezione INFN, Trieste, Italy
\item \Idef{org104}Sezione INFN, Turin, Italy
\item \Idef{org105}SUBATECH, Ecole des Mines de Nantes, Universit\'{e} de Nantes, CNRS-IN2P3, Nantes, France
\item \Idef{org106}Technical University of Split FESB, Split, Croatia
\item \Idef{org107}The Henryk Niewodniczanski Institute of Nuclear Physics, Polish Academy of Sciences, Cracow, Poland
\item \Idef{org108}The University of Texas at Austin, Physics Department, Austin, TX, United States
\item \Idef{org109}Universidad Aut\'{o}noma de Sinaloa, Culiac\'{a}n, Mexico
\item \Idef{org110}Universidade de S\~{a}o Paulo (USP), S\~{a}o Paulo, Brazil
\item \Idef{org111}Universidade Estadual de Campinas (UNICAMP), Campinas, Brazil
\item \Idef{org112}University of Houston, Houston, Texas, United States
\item \Idef{org113}University of Jyv\"{a}skyl\"{a}, Jyv\"{a}skyl\"{a}, Finland
\item \Idef{org114}University of Tennessee, Knoxville, Tennessee, United States
\item \Idef{org115}University of Tokyo, Tokyo, Japan
\item \Idef{org116}University of Tsukuba, Tsukuba, Japan
\item \Idef{org117}Universit\'{e} de Lyon, Universit\'{e} Lyon 1, CNRS/IN2P3, IPN-Lyon, Villeurbanne, France
\item \Idef{org118}V.~Fock Institute for Physics, St. Petersburg State University, St. Petersburg, Russia
\item \Idef{org119}Variable Energy Cyclotron Centre, Kolkata, India
\item \Idef{org120}Vestfold University College, Tonsberg, Norway
\item \Idef{org121}Warsaw University of Technology, Warsaw, Poland
\item \Idef{org122}Wayne State University, Detroit, Michigan, United States
\item \Idef{org123}Wigner Research Centre for Physics, Hungarian Academy of Sciences, Budapest, Hungary
\item \Idef{org124}Yale University, New Haven, Connecticut, United States
\item \Idef{org125}Yonsei University, Seoul, South Korea
\item \Idef{org126}Zentrum f\"{u}r Technologietransfer und Telekommunikation (ZTT), Fachhochschule Worms, Worms, Germany
\end{Authlist}
\endgroup

\end{document}